\documentstyle[12pt]{article}

\def\journal{\topmargin .3in    \oddsidemargin .5in
        \headheight 0pt \headsep 0pt
        \textwidth 5.625in % 1.2 preprint size  %6.5in
\textheight 8.25in % 1.2 preprint size 9in
        \marginparwidth 1.5in
        \parindent 2em
        \parskip .5ex plus .1ex         \jot = 1.5ex}
\journal

\begin{document}
\begin{titlepage}
\begin{center}
%\date               \hfill   LBNL-    \\
October 7, 1997      \hfill    LBNL-40877\\
Revised December 18, 1997        \hfill hep-ph/9710308

\vskip .5in

{\large \bf Combining real and virtual Higgs boson \\
mass constraints}\footnote
{This work was supported by the Director, Office of Energy
Research, Office of High Energy and Nuclear Physics, Division of High
Energy Physics of the U.S. Department of Energy under Contract
DE-AC03-76SF00098.}

\vskip .5in
%\vskip .2in

Michael S. Chanowitz\footnote{Email: chanowitz@lbl.gov}

%\vskip .1in
\vskip .2in
{\em Theoretical Physics Group\\
     Lawrence Berkeley National Laboratory\\
     University of California\\
     Berkeley, California 94720}
\end{center}

\vskip .25in \begin{abstract} Within the framework of the standard 
model we observe that there is a significant discrepancy between the 
most precise $Z$ boson decay asymmetry measurement and the limit from 
direct searches for Higgs boson production.  Using methods inspired by 
the Particle Data Group we explore the possible effect on fits of the 
Higgs boson mass.  In each case the central value and the 95\% 
confidence level upper limit increase significantly relative to the 
conventional fit.  The results suggest caution in drawing conclusions 
about the Higgs boson mass from the existing data.
\end{abstract} 
\end{titlepage}

%THIS PAGE (PAGE ii) CONTAINS THE LBL DISCLAIMER
%TEXT SHOULD BEGIN ON NEXT PAGE (PAGE 1)
\renewcommand{\thepage}{\roman{page}}
\setcounter{page}{2}
\mbox{ }

\vskip 1in

\begin{center}
{\bf Disclaimer}
\end{center}

\vskip .2in

\begin{scriptsize}
\begin{quotation}
This document was prepared as an account of work sponsored by the United
States Government. While this document is believed to contain correct
 information, neither the United States Government nor any agency
thereof, nor The Regents of the University of California, nor any of their
employees, makes any warranty, express or implied, or assumes any legal
liability or responsibility for the accuracy, completeness, or usefulness
of any information, apparatus, product, or process disclosed, or represents
that its use would not infringe privately owned rights.  Reference herein
to any specific commercial products process, or service by its trade name,
trademark, manufacturer, or otherwise, does not necessarily constitute or
imply its endorsement, recommendation, or favoring by the United States
Government or any agency thereof, or The Regents of the University of
California.  The views and opinions of authors expressed herein do not
necessarily state or reflect those of the United States Government or any
agency thereof, or The Regents of the University of California.
\end{quotation}
\end{scriptsize}

\vskip 2in

\begin{center}
\begin{small}
{\it Lawrence Berkeley National Laboratory is an equal opportunity employer.}
\end{small}
\end{center}

\newpage

\renewcommand{\thepage}{\arabic{page}}
\setcounter{page}{1}
%THIS IS PAGE 1 (INSERT TEXT OF REPORT HERE)
%starthere

{\it \underline {Introduction} } Perhaps the most pressing question in 
particle physics today is the mass scale of the quanta that break 
electroweak symmetry, 
giving mass to the particles in the 
theory, including the quark and lepton constituents of ordinary atomic 
matter.
That scale determines whether the symmetry breaking force is 
weak or strong and it sets the energy scale future accelerators will 
need for detailed studies of the mass-generating mechanism.  In 
general the issue can only be resolved by discovering the symmetry 
breaking quanta at a high energy collider.  However, in particular 
theoretical frameworks, such as for instance the standard model, 
radiative corrections to already measured quantities can be used to 
constrain the mass of the symmetry breaking sector.

Interpreted in the standard model framework, beautiful data from LEP, 
SLAC, and Fermilab appear to favor a light Higgs boson with mass of 
order 100 GeV.\cite{quast} The conclusion emerges from the effect of 
{\em virtual} Higgs bosons, via radiative corrections, on precision 
measurements of the $Z$ and $W$ bosons.  In addition, the four LEP 
experiments have searched for {\em real} Higgs bosons, with negative 
results that when combined are expected to imply a lower limit $m_{H} 
\geq 77$ GeV at 95\% confidence level.\cite{murray} Taken together 
the experiments suggest a window between 80 and a few hundred GeV. The 
purpose of this note is to suggest that the window may in fact be 
substantially larger, in part because of well known inconsistencies 
within the precision data, but more because of equally significant 
inconsistencies between precision data and the direct searches whose 
magnitude has, with some noteworthy 
exceptions,\cite{gurtu,dsw} gone largely unnoticed and/or unremarked.

The problem of how to combine inconsistent data has led to the 
break-up of many beautiful friendships.  The mathematical theory of 
statistics provides no magic bullets and ultimately the discrepancies 
can only be resolved by future experiments.  The Particle Data 
Group\cite{pdg} (PDG) has for many years scaled the uncertainty  
of discrepant results by a factor I will call $S_{\rm PDG}$, 
defined by $S_{\rm {PDG}} = \sqrt{\chi^{2}/(N-1)}$, 
where $N$ is the number of data points being combined.  They scale 
the uncertainty of the combined fit by the factor $S_{\rm PDG}$ if and 
only if $S_{\rm PDG} > 1$.  This is a conservative prescription, which 
amounts to requiring that the fit have a good confidence level, 
ranging from 32\% for $N=2$ to greater than 40\% for larger values 
of $N$.  If the confidence level is already good, the scale factor has 
little effect; it only has a major effect on very discrepant data.  
The PDG argues (see \cite{spdg}) that low confidence level fits occur 
historically at a rate significantly greater than expected by chance, 
that major discrepancies are often, with time, found to result from 
underestimated systematic effects, and that the scaled error provides 
a more cautious interpretation of the data.  A few 
authors\cite{gurtu,jlr,dgps} have applied $S_{\rm PDG}$ to the 
asymmetry measurements of $sin^{2}\theta^{lepton}_{eff}$, as I will 
also do here.

With the top quark mass fixed at the value determined by CDF and D0, 
the most sensitive probe of $m_{H}$ is currently the effective 
leptonic weak interaction mixing angle, 
$sin^{2}\theta^{lepton}_{eff}$, measured in a variety of $Z$ boson 
decay asymmetries.  The extent to which the asymmetries currently 
dominate the estimate of $m_{H}$ can be seen by comparing the 
conventional (unscaled) fit to the seven asymmetry measurements, 
$m_{H}= 104 \pm^{110}_{54}$GeV, with the LEP electroweak working 
group\cite{lepewwg} global fit to all data, $m_{H}= 115 
\pm^{116}_{66}$GeV. Because it more than suffices for the purposes of 
this paper, the analysis that follows is based on the asymmetry 
measurements alone.  The PDG scale factor then increases the 
uncertainty but not the central value of the combined fit for 
$sin^{2}\theta^{lepton}_{eff}$ and $m_{H}$.\footnote{If instead the 
rescaled fit to $sin^{2}\theta^{lepton}_{eff}$ is included in a global 
fit to all $Z$ decay data, as in Gurtu's analysis\cite {gurtu}, the 
rescaling does affect the central value of $m_{H}$.}

The focus of this paper is on the discrepancies between precision 
measurements and the limit from the direct searches, which will be 
addressed by a method analogous to the PDG scale factor.  Like the PDG 
prescription, the idea is to scale the error so that the precision 
measurement has a significant probability $P$ to be consistent with 
the direct search limit.  I consider $P=0.32$, corresponding to the 
PDG's choice, as well as larger and smaller values.  To account for 
uncertainty in the search limits, which may also be subject to unknown 
systematic errors, I consider a range of different lower limits on 
$m_{H}$, from a very conservative 50 GeV to a futuristic 90 GeV. In 
this approach both the central value and the uncertainty of the fit 
are affected.  In addition I present fits using two other methods 
discussed by the PDG. 

By giving full weight to a measurement that is in serious conflict 
with the direct search lower limit, the conventional method risks 
underestimating $m_{H}$.  The alternative methods considered here 
provide a more conservative estimate of the upper limit on $m_{H}$ but 
risk skewing the fit to large $m_{H}$.  Taken together the results 
strongly suggest caution in drawing conclusions from the precision 
data about the value of the Higgs boson mass.

{\it \underline {The precision data}} 
The relevant values of $sin^{2}\theta^{lepton}_{eff}$ and the quoted 
experimental uncertainties are shown in table 1, from the preliminary 
values presented at the 1997 summer conferences.\cite{quast} For 
each value the table displays the corresponding value of $m_{H}$ and 
the 95\% CL upper ($m_{95}^{<}$) and lower ($m_{95}^{>}$) bounds (that 
is, the symmetric 90\% confidence intervals).  Also indicated is the 
probability for $m_{H}$ to lie below 77 GeV. Gaussian distributions 
are assumed for $sin^{2}\theta^{lepton}_{eff}$ and $log(m_{H})$.

The values of $m_{H}$ are from the state of the art $\overline{MS}$ 
computation of reference \cite{dgps} --- see also \cite{dgs}.  To 
obtain the confidence intervals and probabilities the parametric error 
is combined in quadrature with the experimental errors.  The 
parametric error is equivalent to $\pm 0.00030$ uncertainty in 
$sin^{2}\theta^{lepton}_{eff}$ ---- see \cite{dgps,dgs}.  It is 
dominated by roughly equal contributions from the uncertainties in the 
top quark mass, $m_{t}=175\pm 6$ GeV, and the fine structure constant 
at the $Z$ mass, $\alpha^{-1}(m_{Z})= 128.896 \pm 0.090$, in addition 
to other much smaller contributions, including $\Delta \alpha_{\rm 
QCD}(m_{Z})$ and uncomputed higher order corrections.\footnote{There 
are also negligible extrapolation errors from reference  
\cite{dgps}, equivalent to $\leq 0.00003$ in 
$sin^{2}\theta^{lepton}_{eff}$ for $75 < m_{H} < 600$ GeV. 
Even outside this range they have no real effect on the 
analysis, since the confidence levels and scale factors only depend on 
the relationship between $sin^{2}\theta^{lepton}_{eff}$ and $m_{H}$ 
for $m_{H}=m_{H}^{\rm {limit}}$. The worst case is then 
$m_{H}^{\rm {limit}}= 50$ GeV, close enough for any additional 
error to be negligible.  The very large values of $m_{95}^{>}$ in the 
tables could be affected but they have no precise significance in any 
case.}

The six LEP measurements in table 1 are each combined from the four 
LEP experiments, and in each case the combined fit has a good 
confidence level.  The conventional maximum likelihood fit for the LEP 
measurements is shown in the first row of table 2.  The chi-squared 
per degree of freedom is $\chi^{2}/N-1=4.4/5$ corresponding to a 
robust 0.5 confidence level. The central value is $m_{H}=240$ GeV 
and the 95\% CL upper limit is 860 GeV. There is no entry for the
$S_{\rm PDG}$ fit since $S_{\rm PDG} < 1$.

Combining all seven measurements (the conventional LEP + SLC fit in 
table 2) the central value decreases to 100 GeV and the 95\% CL upper
limit falls to 310 GeV, demonstrating the effect of the high precision 
and lower $sin^{2}\theta^{lepton}_{eff}$ from $A_{LR}$.  The 
chi-squared per degree of freedom now rises to 12.5/6, with a marginal 
confidence level of 0.05.  The PDG scale factor is then $S_{\rm 
PDG}=1.45$.  Using it, the combined uncertainty of the fit increases 
from $\pm 0.00023$ to $\pm 0.00033$ and the 95\% CL upper limit on 
$m_{H}$ increases modestly to 420 GeV.

{\it \underline {\ldots and the direct search limits}} In addition to 
discrepancies among the measurements of 
$sin^{2}\theta^{lepton}_{eff}$, which are problematic whether we 
assume the standard model or not, table 1 also reveals a second 
discrepancy that occurs specifically within the standard model 
framework.  The most precise measurement, $A_{LR}$, implies a 95\% CL 
{\em upper} limit on $m_{H}$ of 77 GeV, while the direct searches at 
LEP are expected to provide a combined 95\% {\em lower} limit also at 
77 GeV.\cite{murray} (The individual 95\% CL limits quoted by the four
experiments range from 66 to 71 GeV.\cite{murray}) The third most 
precise measurement, $A_{FB}^{l}$, also has significant weight (71\%) 
below the direct search limit.\footnote{$A_{LR}$ and $A_{FB}^{l}$ are 
also the only measurements with $m_{95}^{>}$ 
%in table 1 that imply 95\% upper limits 
below the TeV scale.}

This raises a difficult question: within the standard model framework 
what role if any should the direct search limits play in extracting 
the implications of the precision data?  There is no single ``right'' 
answer to the question.  A maximum likelihood fit including both the 
precision data and the direct search data would replicate the 
conventional fit if the central value lies above the lower limit, 
$m_H^{\rm limit}$, from the direct searches.  That is a defensible 
interpretation, since if the true value of $m_H$ were near $m_H^{\rm 
limit}$ we would expect values of $m_H$ obtained from measurements of 
$sin^{2}\theta^{lepton}_{eff}$ to lie both above and below $m_H^{\rm 
limit}$.  By underweighting downward fluctuations while leaving upward 
fluctuations at their full weight, we risk skewing the fit upward.  
Mindful of this risk, it is still instructive to explore the 
sensitivity of the fit to the weight ascribed to measurements that are 
{\em individually in significant contradiction} with the direct search 
limit.

Clearly the direct search results are not irrelevant.  If, for 
instance, the only information available were the direct search limits 
and the $A_{LR}$ measurement, we would conclude that the standard 
model is excluded at 90\% CL.\footnote{I
thank Lawrence Hall for this perspective.} Theorists 
would have flooded the Los Alamos server with papers on the death 
of the standard model and the birth of new theories W,X,Y,Z...  In the 
actual situation the $A_{LR}$ measurement causes the fit to $m_{H}$ to 
shift by more than a factor two, from 240 GeV to 100 GeV, and the 95\% 
upper limit to fall from the TeV scale to $\simeq 340$ GeV. It is 
fully weighted in the conventional standard model fit despite a 
significant contradiction with the standard model.

If the discrepancy were even greater --- say, for instance, a precision 
measurement implying $m_{H}=10$ MeV with a 99.99\% CL upper
limit at 77 GeV --- the clear response would be to omit that 
measurement from a standard model fit, although it could still be 
considered in a broader framework encompassing the possibility of new 
physics.\footnote {In fact, parity violation in atomic Cesium 
currently implies $m_{H}\sim 11 $ MeV ({\it M}eV is not a 
typographical error) though only $1.2\sigma$ from 77 
GeV.\cite{deandrea} Its weight in the combined fit would be 
negligible.} On the other hand, $A_{FB}^{l}$, with 31\%
probability to be consistent with a 95\% lower
limit at 77 GeV, would surely be retained.  The question is how to 
resolve the intermediate cases in which the discrepancy is significant 
but not so significant that the data should clearly be excluded.

Consider a prescription, analogous to the PDG scale factor, that 
interpolates smoothly between the extremes.  Imagine a measurement $x$ 
with experimental error $\delta_{E}$ and a quantity $y$ that is 
related to $x$ with an uncertainty $x \pm \delta_{P}$ (the parametric 
error).  Suppose there is an upper limit on $y$ at 
$y=y_{0}$ that translates to an upper limit on $x$ at 
$x_{0}\pm\delta_{P}$, such that the measurement $x$ falls below the 
implied limit, $x < x_{0}$.  The discrepancy between the measurement 
and the limit is then characterized by a Gaussian distribution 
centered at $x$ with standard deviation $\sigma = \sqrt{\delta_{E}^{2} 
+ \delta_{P}^{2}}$, with a computable probability $P$ for $x>x_{0}$.  
If $P$ is less than a chosen minimal confidence level $P_{VR}$ ($VR$ 
for ``virtual-real''), then $\delta_{E}$ is scaled by a factor $S_{VR}$ 
chosen so that the Gaussian centered at $x$ with standard deviation 
$\sigma' = \sqrt{(S_{VR}\delta_{E})^{2} + \delta_{P}^{2}}$ has 
probability $P_{VR}$ for $x>x_{0}$.  If $x_{0} - x$ is small enough, 
the scale factor has little or no effect.  If $x$ is many $\sigma$ 
below $x_{0}$, $S_{VR}$ will be large and the data point $x$ will have 
reduced weight in a combined fit with other data.  Intermediate cases 
will interpolate smoothly between the two extremes, depending  
on the values of $x-x_{0}$, $\sigma$, and $P_{VR}$.

The value of $P_{VR}$ is of course arbitrary.  One plausible choice is 
$P_{VR}=0.32$ (or 0.3173 to the cognoscenti), since that is the 
confidence level implicit in the PDG scale factor for $N=2$.  A 
plausible choice for the lower limit on $m_{H}$ is $m_{H}^{{\rm 
limit}}= 77$ GeV. The resulting fits are shown in table 2.  The fit to 
the LEP data is affected only modestly, with an increase of $\sim10\%$ 
in $m_{H}$.  For the LEP + SLC fit, the central value of $m_{H}$ and 
the 90\% confidence interval increase significantly, to the values of 
the LEP fit.  The scale factors are $S_{VR}(A_{FB}^{l})=1.2$ and 
$S_{VR}(A_{LR})=4.1$.

Table 3 displays the results of varying $P_{VR}=0.20,\ 0.32,\ 0.40$ 
and $m_{H}^{{\rm limit}}= 50,\ 60,\ 70,\ 80,\ 90$ GeV. Though not 
reported by the experimental groups, the confidence levels at 50 and 
60 GeV are probably much tighter than 97\%, both because
 the LEP II cross sections increase for smaller $m_{H}$ and because 
 the LEP I data contributes to the confidence level at those masses 
 ---- the 95\% CL lower bound from LEP I alone is $m_{H}>66$
 GeV. The value 80 GeV is close to the presently projected 95\% 
 combined limit of the four LEP experiments, while 90 GeV is the 
 anticipated limit if no discovery emerges from currently planned LEP 
 II running.  Vacant entries in table 3 indicate that the fits are 
 unmodified, $S_{VR} \leq 1$, and that the conventional fit, line 1 of 
 table 2, applies.  For the LEP data, the Higgs mass scale varies by 
 no more than a factor 1.5 from the conventional fit over the entire 
 range of table 3.  For the LEP + SLC data, the difference is a factor 
 1.5 for $(P_{VR}, m_{H}^{{\rm limit}})=(0.20, 50$ GeV) and becomes as 
 large as a factor 4.
 
 {\it \underline {Other methods}} 
In this section I will briefly present results using
methods discussed by the PDG\cite{pdgstat} for combining 
measurements that conflict with a limit. They are no less 
arbitrary than the $S_{VR}$ scale factor method discussed above.

Consider a collection of measurements $x_{i} \pm \delta_{E,i}$, 
$i=1,\ldots , N$, some of which are nominally inconsistent with an 
upper limit at $x_{0}$.  The ``Bayesian'' method is to combine all 
data points in the conventional way and to multiply the combined 
Gaussian distribution by a step function $C\theta(x-x_{0})$, so that 
the distribution vanishes below $x_{0}$.  $C$ is a normalization 
factor to guarantee total unit probability.  I have modified the usual 
Bayesian method to account for the fact that the lower limit is not 
absolute but has 95\% confidence, by choosing $C$ to give the 
distribution probability 0.95 for $x>x_{0}$.  The 50\%- and 95\%-tiles 
of the resulting distributions are shown in table 2 for $m_{H}^{{\rm 
limit}}=77$ GeV. 

Three ``frequentist'' prescriptions are also discussed by the PDG, of 
which two are considered here.  For $x_{i}<x_{0}$ one prescription 
assigns $x_{i}\pm\delta_{E,i} \rightarrow x_{0}\pm\delta_{E,i}$ when 
the limit $x_{0}$ is known exactly.  Including the parametric 
uncertainty, I modify this to $x_{i}\pm\delta_{E,i} \rightarrow 
x_{0}\pm\sigma_{i}$ where $\sigma_{i}=\sqrt{\delta_{E,i}^{2} + 
\delta_{P}^{2}}$.  The readjusted points are then combined as usual 
(including $S_{\rm PDG}$ if applicable) with the other measurements.  
An extremely conservative variation, intended only to obtain the 95\% 
CL upper limit, replaces $x _{i} \rightarrow {\rm min}(x _{i}, x_{0} + 
1.64\sigma_{i})$, so that 95\% of the probability distribution for 
each measurement is above the limit $x_{0}$.  The results are 
illustrated in table 2 for $m_{H}^{{\rm limit}}=77$ GeV.

{\it \underline {Conclusion}} Several related points are deferred to 
a more detailed report, including the following: (1) There are 
indications, depending on how the data is grouped, that the 
confidence level of the conventional fit may be even less than 
0.05.  Dependence on how the data is grouped reflects the uncertainty 
in the confidence level for fits of small data samples.  (2) Though 
definitive conclusions can only come from the experimental groups, 
two estimates suggest that improved $b$-tagging methods are not 
likely to cause a big shift in $A_{FB}^{b}$.  (3) The $W$ boson mass 
measurement currently has a sensitivity to $m_{H}$ at roughly the 
middle of the pack of the asymmetry measurements.  It does not 
qualitatively alter the conclusions.

In summary, the $A_{LR}$ measurement is inconsistent at 95\% CL both 
with the LEP asymmetry measurements and, in the standard model, with 
the Higgs boson search limits, while its precision causes it to have a 
profound effect on the combined standard model fit.  The conflict with 
the search limits may diminish or disappear if there is new physics 
outside the standard model framework but is necessarily germane to a 
standard model fit.  The analysis presented here is meant as a warning 
signal, a yellow if not a red flag, suggesting caution in drawing 
conclusions from the precision data about the mass of the standard 
model Higgs boson.  Applying methods inspired by the Particle Data 
Group to these discrepancies, we find that the central value of 
$m_{H}$ increases by factors from $\sim 1.5$ to $\sim 3$ while the 
95\% CL upper limit increases toward the TeV scale.  Only future 
experimental results can resolve the discrepancies in the present 
experimental situation.

\noindent{\bf Acknowledgements:} I wish to thank Michael Barnett, 
Robert Cahn, Donald Groom, Lawrence Hall, and Gerry Lynch for useful 
discussions.  This work was supported by the Director, Office of 
Energy Research, Office of High Energy and Nuclear Physics, Division 
of High Energy Physics of the U.S. Department of Energy under 
Contract DE-AC03-76SF00098.

\newpage
\vskip 0.5in
\noindent {\bf Tables}
\vskip 0.2in

Table 1.  Values for $sin^{2}\theta^{lepton}_{eff}$ from asymmetry 
measurements\cite{quast} with $1\sigma$ experimental errors.  The 
corresponding Higgs boson masses, the 95\% CL upper and lower limits,  
and the confidence level for $m_{H} < 77$ GeV are given for each 
measurement. 

\begin{center}
\vskip 20pt
\begin{tabular}{c|cccc}
	& $\bf{sin^{2}\theta^{lepton}_{eff}}$ \bf{(1}$\bf{\sigma}$)& 
	$\bf{m_{H}}$ \bf{(GeV)}& $\bf{m_{95}^{>},\ m_{95}^{<}}$ & 
	$\bf{P(<77 {\rm GeV})}$ \\
\hline
\hline
$A_{LR}$ & 0.23055 (41) & 16 & 3, 80 & 0.95 \\
\hline
$A_{FB}^{b}$ & 0.23236 (43) & 520 & 100, 2700 & 0.03 \\
\hline
$A_{FB}^{l}$ & 0.23102 (56) & 40 & 5, 290 & 0.71 \\
\hline
$A_{\tau}$ & 0.23228 (81) & 440 & 30, 6700 & 0.14 \\
\hline
$A_{e}$ & 0.23243 (93) & 590 & 28, 13000 & 0.14 \\
\hline
$Q_{FB}$ & 0.23220 (100) & 380 & 14, 10000 & 0.21 \\
\hline
$A_{FB}^{c}$ & 0.23140 (111) & 83 & 2, 3000 & 0.49 \\
\hline
\hline 
\end{tabular}
\end{center}

\newpage
Table 2. Fits to the LEP and LEP + SLC data as described in the text. 
The Bayesian and frequentist fits assume $m_{H}^{{\rm limit}}=77$ 
GeV. 
\begin{center}
\vskip 20pt
\begin{tabular}{c|c|ccc}
\bf{Data set}&\bf{Fit}& $\bf{sin^{2}\theta^{lepton}_{eff}}$\bf 
{(1}$\bf {\sigma }$)&$\bf{m_{H}}$ \bf {(GeV)}& $\bf 
{m_{95}^{<},\ m_{95}^{>}}$ \\
&&&&\\
\hline
\hline
&&&&\\
&Conventional & 0.23196 (28) & 240 & 67, 860 \\
%\hline
%&&&&\\
&$S_{PDG}$ & & & \\
%\hline
%&&&&\\
{\bf LEP}&$S_{VR}(0.32, 77)\ {\rm GeV})$ & 0.23203(29) & 270 & 
74,1000 \\
%\hline
%&&&&\\
&Bayes & 0.23197 & 250 & $< 880$ (95\%)\\ 
%\hline
%&&&&\\
&Frequentist (1) & 0.23204(28) & 280 & 78,1000 \\
%\hline
%&&&&\\
&Frequentist (2) & $< 0.23304$ (95\%) & & $²1900$ (95\%) \\
&&&&\\
\hline
\hline
&&&&\\
&Conventional& 0.23152 (23) & 100 & 32, 340 \\
%\hline
%&&&&\\
&$S_{PDG}$ & 0.23152 (33) & 100 & 26, 420 \\
%\hline
%&&&&\\
{\bf LEP + SLC } &$S_{VR}(0.32, 77)\ {\rm GeV})$& 0.23198 (28) 
& 250 & 69,920 \\
%\hline
%&&&&\\
&Bayes & 0.23171 & 150 & $<500$ (95\%) \\
%\hline
%&&&&\\
&Frequentist (1)& 0.23183 (23) & 190 & 58,610 \\
%\hline
%&&&&\\
&Frequentist (2)& $<0.23284$ (95\%) & & $<1300$ (95\%) \\
&&&&\\
\hline
\hline

\end{tabular}
\end{center}

\newpage 

Table 3.  Fits to LEP and LEP + SLC data using the $S_{VR}$ scale 
factor with various values of $P_{VR}$ and $m_{H}^{\rm limit}$.  Each 
entry displays the central value of $m_{H}$ and $m_{95}^{<},\ 
m_{95}^{>}$, the 95\% CL lower and upper limits, 
in GeV. Empty entries indicate that no measurement is far enough 
below threshold to be modified by the scale factor and that the 
conventional fit of table 2 applies.

\begin{center}
\vskip 20pt
\begin{tabular}{c|ccc|ccc}
&&{\bf LEP} &&&{\bf LEP + SLC}  & \\ 
\hline 
$m_{H}^{\rm limit}$& $P_{VR}=0.20$ & 0.32 & 0.40 & 0.20 & 0.32 & 0.40 \\
\hline
\hline
50 & &  & & 160 & 210 & 230 \\
 & & & & 41,590 & 59,740 & 65, 830 \\
\hline
60 &  & & 310 & 170 & 220 & 300 \\
& & & 82, 1100 & 47, 650 & 61,770 & 80, 1100 \\
\hline
70 & & & 360 & 190 & 220 & 350 \\
& & & 94, 1400 & 51, 680 & 62,790 & 91, 1300 \\
\hline
80 & & 290 & 380 & 190 & 260 & 370 \\
& & 77,1100 & 100, 1500 & 54,700 & 72,970 & 96,1400 \\
\hline
90 & & 320 & 390 & 200 & 300 & 380 \\
& & 85,1200 & 100, 1500 & 56, 710 & 79,1100 & 100, 1500 \\
\hline
\hline 
\end{tabular}
\end{center}

\end{document}